\documentstyle[aps,prd,epsf]{revtex}

\newcommand{\be}{\begin{equation}}
\newcommand{\ee}{\end{equation}}
\newcommand{\ben}{\begin{eqnarray}
\displaystyle}
\newcommand{\een}{\end{eqnarray}}

\newcommand{\p}{\partial}
\newcommand{\na}{\nabla}

\newcommand{\tg}{{\tilde g}}

\newcommand{\trho}{{\tilde \rho}}
\newcommand{\th}{{\tilde h}}

\newcommand{\tR}{{\tilde R}}

\newcommand{\ep}{\epsilon}

\newcommand{\ga}{\gamma}

\begin{document}

\title{Uniqueness Theorem for Static Black Hole Solutions of $\sigma$-models
 in Higher Dimensions}

\author{Marek Rogatko}

\address{Institute of Physics \protect \\
Maria Curie-Sklodowska University \protect \\
20-031 Lublin, pl.Marii Curie-Sklodowskiej 1, Poland \protect \\
rogat@tytan.umcs.lublin.pl \protect \\
rogat@kft.umcs.lublin.pl}

\date{\today}

\maketitle

\begin{abstract}
We prove the uniqueness theorem for self-gravitating non-linear 
$\sigma$-models in higher dimensional spacetime. Applying the positive 
mass theorem we show that Schwarzschild-Tagherlini spacetime is the only
maximally extended, static asymptotically flat solution with
non-rotating regular event horizon with a constant mapping.
\end{abstract}

\pacs{04.20.Cv}

\baselineskip=18pt
Nowadays, much effort is being devoted to the study of mathematical topics
related to the black hole equilibrium states. The pioneering investigations 
in this field were attributed to Israel \cite{isr},
M\"uller zum Hagen {\it et al.} \cite{mil73} and Robinson \cite{rob77}.
The alternative approach to the problem of the uniqueness of black hole
solutions
was proposed by Bunting and Masood-ul-Alam \cite{bun}
and then strengthened to the Einstein-Maxwell (EM) black holes
\cite{ru,ma1}. Heusler \cite{he1} comprised the 
magnetically charged RN solution and static Einstein-${\sigma}$-model case
\cite{he93}.
Recently the classification of static of 
vacuum black holes was finished \cite{chr99a}.
The condition of non-degeneracy of the 
event horizon was removed and it was shown that Schwarzschild black hole exhausted
the family of all appropriately regular black hole spacetimes.
In \cite{chr99b} it was
revealed 
that RN solution comprised the family of regular
black hole spacetimes under the restrictive condition that all
degenerate components of black hole horizon carried a charge of the same sign.
\par
The uniqueness of stationary
and axisymmetric black hole spacetimes being the solution of vacuum 
Einstein equations were achieved in \cite{car73,car87,rob75,wal71},
while 
the systematic way of obtaining the desire results in
electromagnetic case was conceived by Mazur 
\cite{maz} and Bunting \cite{bun}.
For a review of the uniqueness of black hole
solutions story see \cite{book} and references therein. 
\par
Nowadays, there has been an active period of constructing black hole solutions
in the string theories (see \cite{you97} and references therein). 
In the low-energy string theory (Einstein-Maxwell-dilaton gravity) the uniqueness
of the black holes was established in \cite{mas93,gur95,mar01}.
The uniqueness theorem for static dilaton $U(1)^2$
black holes being the solution of $N = 4, d = 4$ supergravity
was provided in \cite{rog99} while the proof of the uniqueness theorem for
static $U(1)^{N}$ dilaton black holes was established in \cite{rog02}.
\par
The development of string theory also trigger the interests in higher
dimensional black hole solutions. 
Another theoretical challenge is the possibility that the weak scale is the fundamental
scale of nature and the Planck scale is to be derived from it \cite{gva}.
The development of the so-called
TeV gravity attract attention to higher dimensional black hole 
which may be produced in high energy experiments \cite{gid}. Thus in order to predict
the phenomenological results one should pay more attention to properties of
higher dimensional black holes.
\par
The five-dimensional stationary vacuum black hole are not unique, Myers-Perry
\cite{mye86} solution generalizes the Kerr solution to arbitrary dimension, while Emparan
{\it et al.} \cite{emp02} presents a counterexample. They found a five-dimensional
rotating black hole ring solution with the same angular momentum and mass
but the horizon of which was homeomorphic to $S^{2} \times S^{1}$.
On the other hand, in static spacetime the uniqueness theorem for $n$-dimensional
Schwarzschild-Tangherlini
black hole was given by
Gibbons {\it et al.} \cite{gib02} and
the proof of the uniqueness theorem for asymptotically flat
static $n$-dimensional
charged dilaton black hole was given in \cite{gib02b}.\\
In this issue we shall consider the simplest 
generalization of the $n$-dimensional static vacuum
solution, namely
we shall consider the static $n$-dimensional black hole solution of
Einstein equations coupled to $n$-dimensional $\sigma$-model with
harmonic action.
One shall try to proof the uniqueness theorem for static
$n$-dimensional black hole solutions in the underlying theory.
In our proof we use the Bunting approach and in the last part
conclusions given in \cite{gib02}.
The action describing $n$-dimensional self-gravitating
$\sigma$-model yield:
\be
I = \int d^n x \sqrt{-g} \bigg[ {}^{(n)}R - 
{1 \over 2}G_{AB}(\phi(x))\phi^{A}_{,\mu}\phi^{B}_{,\nu}
\bigg],
\label{act}
\ee
while the energy-momentum tensor for $\sigma$-model is as follows:
\be
T_{\mu \nu}(\phi) = G_{AB}(\phi(x))\phi^{A}_{,\mu}\phi^{B}_{,\nu}
- {1 \over 2}G_{AB}(\phi(x))\phi^{A}_{,\xi}\phi^{B ,\xi} g_{\mu \nu}.
\ee
The equation of motion derived by means of the variational principle yield
\ben
\na_{\mu}\na^{\mu} \phi^{A} + 
\Gamma^{A}_{BC}\phi^{B}_{,\mu}\phi^{C ,\mu} &=& 0, \\ 
{}^{(n)}G_{\mu \nu} &=& T_{\mu \nu}(\phi).
\een
\par
In our considerations we shall assume that one has to do with a non-rotating
black hole with strictly stationary domain of outer communication $<<J>>$
defined as in \cite{book}. By this statement one understands that the null generators
of the event horizon will coincide with a Killing vector field $k_{\mu}$ and
Killing vector field is timelike in all domain of outer communication.
The non-degenerate Killing horizon will be obtained if we consider the
existence of a compact bifurcation surface in the spacetime under 
consideration. It turned out that if we deal with an asymptotically flat,
globally hyperbolic spacetime with a compact bifurcation surface and strictly
stationary, simply connected domain of outer communication $<<J>>$, i.e.,
$V = - k_{\mu}k^{\mu} \ge 0$, then $<<J>>$ is static iff the following is
satisfied \cite{book}:
\be
\omega_{\alpha} = 0 \Leftrightarrow k^{\ga}R_{\ga [ \alpha}k_{\rho ]} = 0,
\label{rrr}
\ee
where we have introduced the rotation vector 
$\omega_{\alpha} = {1 \over 2} 
\ep_{\alpha \beta \ga \delta} k^{\beta} \na^{\ga}k^{\delta}$.
We shall also require that the field $\phi$ obey the stationarity condition
with respect to the Killing vector field $k_{\mu}$, namely
\be
{\cal L}_{k} \phi = 0.
\ee
Having the above in mind and using equation of motion we get that 
$R_{\alpha \beta}k^{\alpha} = 0$. Thus, the rotation vector 
defined by relation (\ref{rrr}) is equal to zero.
\par
In the domain of outer communication $<<J>>$
the metric of $n$-dimensional static spacetime has the following form:
\be
ds^2 = - V^2 dt^2 + g_{i j}dx^{i}dx^{j},
\ee
where $V$ and $g_{i j}$
are independent of the $t$-coordinate as the quantities
of the hypersurface $\Sigma$ of constant $t$. Let us assume further
that in asymptotically flat spacetime there is a standard coordinates
system in which we have the usual asymptotic expansion
\be
V = 1 - {C \over r^{n - 3}} + {\cal O}\bigg( {1 \over  r^{n - 2}} \bigg),
\ee
and
\be
g_{ij} = \bigg( 1 + {2 \over n - 3}{C \over r^{n - 3}} \bigg)+
{\cal O} \bigg( {1 \over r^{n - 2}} \bigg),
\ee
where $C$ is the ADM mass seen by the observer from the infinity, while
$r^2 = x_{i}x^{i}$.
The static Einstein equations 
and the equations for self-gravitating harmonic mappings
on $\big( \Sigma, g_{ij} \big)$ imply
\ben
{}^{(n - 1)}R_{ij}(g) - {1 \over V}{}^{(g)}\na_{i} {}^{(g)}\na_{j} V &=&
G_{AB}(\phi(x))\phi^{A}_{,i}\phi^{B }_{,j}, \\
{}^{(n - 1)}R(g) - {2 \over V}{}^{(g)}\na_{i} {}^{(g)}\na^{i} V &=&
G_{AB}(\phi(x))\phi^{A}_{,i}\phi^{B ,i}, \\
V{}^{(g)}\na_{i} {}^{(g)}\na^{i} V &=& 0,
\een 
where ${}^{(g)}\na$ is the covariant derivative 
with respect to 
$g_{ij}$
while ${}^{(n - 1)}R_{ij}(g)$ is the Ricci tensor defined on 
the hypersurface $\Sigma$.
In order to apply the positive mass theorem in our uniqueness proof
one should paste together two copies of $\Sigma$, i.e.,
$\Sigma_{+}$ and $\Sigma_{-}$ with boundaries ${\cal H}_{+}$ and
${\cal H}_{-}$ and metrics given by
\be
\tg^{\pm}_{ij} = {\Omega_{\pm}}^{2} g_{ij},
\ee
and the conformal factors $\Omega_{\pm}$ in the forms as follows:
\be
\Omega_{\pm} = \bigg( {1 \pm V \over 2} \bigg)^{2 \over n - 3}.
\ee
Pasting $\Sigma_{+}$ and $\Sigma_{-}$ across the horizon for which $V = 0$
we can construct a complete regular hypersurface $(\Sigma_{+} \cup \Sigma_{-}
\cup {\cal P}, \tg_{ij})$, where ${\cal P}$ is the point at infinity.
Moreover, one can show that $\Omega_{\pm}$ vanishes nowhere in $\Sigma$
and $\tg_{ij}$ is extendible to $\Sigma_{-} \cup {\cal P}$.\\
Using the asymptotical 
expansion for $g_{ij}$ and $V$, one gets immediately \cite{gib02}
\be
\tg^{+}_{ij} = \delta_{ij} + {\cal O} \bigg( {1 \over r^{n - 2}} \bigg),
\ee
and for the behaviour on $\Sigma^{-}$ hypersurface we obtain
\be
\tg^{-}_{ij} dx^{i} dx^{j} = {{(C / 2)^{4 \over n - 3}} \over r^4}
\big( dr^2 + r^2 d\Omega^{2}_{n - 3} \big) + {\cal O} \bigg(
{1 \over r^5} \bigg).
\ee
The asymptotic behaviour of $\tg^{+}_{ij}$ provides that $\Sigma_{+}$
is asymptotically flat and has vanishing mass.\\
The next task is to show that the Ricci scalar $\tR_{ij}(\tg)$ is non-negative.
The conformally scaled Ricci curvature has the form
\be
{}^{(n - 1)}\tR (\tg) = {
{}^{(n - 1)}R(g) \over \Omega^2} - {4 (n - 2)\over \Omega^2 (n - 3)}
{{}^{(g)}\na_{i} {}^{(g)}\na^{i} V \over (1 + V)}.
\ee
Using equations of motion for the underlying theory one gets
\be
{}^{(n - 1)}\tR (\tg) = {1 \over \Omega^2} G_{AB}(\phi(x))\phi^{A}_{,i}\phi^{B ,i}.
\label{cr}
\ee
From relation (\ref{cr}) one can see that the Ricci tensor 
${}^{(n - 1)}\tR (\tg)$ on the hypersurface $\Sigma$ is manifestly
non-negative. Equation (\ref{cr})
depicts one more fact, namely that the scalar field $\phi$ should be constant.
\par
As a consequence the manifold $(\Sigma_{+} \cup \Sigma_{-}
\cup {\cal P}, \tg_{ij})$ satisfies the conditions to apply
the positive mass theorem \cite{posth}. This implies in turn that
$(\Sigma_{+} \cup \Sigma_{-}
\cup {\cal P}, \tg_{ij})$ is isometric to flat manifold.
Thus
we can rewrite $g_{ij}$ in a 
conformally flat form \cite{gib02}
\be
g_{ij} = {\cal U}^{4 \over n-3} \delta_{ij},
\label{gg}
\ee
where ${\cal U} = {2 \over 1 + V}$. \\
The rest of the uniqueness proof proceeds exactly as in the vacuum case so for 
the readers' convenience we shall outline the main points of it. Namely,
using relation (\ref{gg}) one can show
that the Einstein-$\sigma$ model equations of motion reduces
to the Laplace equation on the $(n - 1)$ Euclidean manifold, namely
\be
\na_{i}\na^{i}{\cal U} = 0,
\ee
where $\na$ is the connection on a flat manifold. 
Applying the expression for the base space
\be 
\delta_{ij} dx^{i}dx^{j} = \trho^{2} d{\cal U}^2 + \th_{AB}dx^{a}dx^{B},
\ee
and having in mind that the event horizon is located at ${\cal U} = 2$,
one can show that the embedding of $\cal H$ into the Euclidean
$(n-1)$ space is totally umbilical \cite{kob69}. This embedding must be 
hyperspherical, i.e., each of the connected components of the horizon $\cal H$
is a geometric sphere with a certain radius determined by the value of
$\rho \mid_{\cal H}$,
where $\rho$ is the coordinate which can be introduced on $\Sigma$
as follows:
$$g_{ij}dx^{i}dx^{j} = \rho^2 dV^2 + h_{AB}dx^{A}dx^{B}.$$
It turns out that one can always locate one 
connected component of the horizon at $r = r_{0}$ surface without loss of generality.
Thus we have a boundary value problem for the Laplace equation 
on the base space $\Omega = E^{n-1}/B^{n-1}$ with the
Dirichlet boundary condition ${\cal U} \mid_{\cal H} = 2$ and the asymptotic
decay condition ${\cal U} = 1 + {\cal O} \bigg( {1 \over r^{n-3}} \bigg)$.
Let ${\cal U}_{1}$ and ${\cal U}_{2}$ be two solutions of the boundary value problem.
Using the Green identity and integrating over the volume element
one gets
\be
\bigg( \int_{r \rightarrow \infty} - \int_{\cal H} \bigg) 
\bigg( {\cal U}_{1} - {\cal U}_{2} \bigg) {\p \over \p r}
\bigg( {\cal U}_{1} - {\cal U}_{2} \bigg) dS = \int_{\Omega}
\mid \na \bigg( {\cal U}_{1} - {\cal U}_{2} \bigg) \mid^{2} d\Omega.
\ee
Then, having in mind the boundary condition the left-hand side vanishes, and one has
that two solutions must be identical.\\
In conclusion we obtain the following theorem:\\
Let $\phi$ be a stationary mapping with harmonic action. 
The only black hole solution with regular, non-rotating 
event horizon in the asymptotically flat, strictly stationary domain of outer
communication is Schwarzschild-Tangherlini
black hole solution with
a constant mapping $\phi$.



\vspace{1cm}
\noindent
{\bf Acknowledgements:}\\
MR was supported in part by KBN grant 5 P03B 009 21.



\end{document}